\documentclass{article}
\font\sqi=cmssq8
\def\DR{\rm I\kern-1.45pt\rm R}
\def\DC{\kern2pt {\hbox{\sqi I}}\kern-4.2pt\rm C}
\def\DH{\rm I\kern-1.5pt\rm H\kern-1.5pt\rm I}

\input epsf

\def\be{\begin{equation}}
\def\ee{\end{equation}}
\def\arr{\begin{array}{rll}}
\def\ea{\end{array}}
\def\bea{\begin{eqnarray}}
\def\eea{\end{eqnarray}}
\def\cN{{\cal{N}}}

\def\N2{$N{=}2$}

\def\>{\rangle}
\def\<{\langle}
\def\+{\dagger}
\def\={\ =\ }

\usepackage{amscd,amsmath,amssymb}
\begin{document}
\begin{center}
{\bf \Large Lobachevsky geometry of (super)conformal mechanics}
\vspace{0.5 cm}

{\large Tigran Hakobyan$^{1,3}$ and  Armen Nersessian$^{1,2}$}
\end{center}
 ${}^{1}$   {\sl Yerevan State University,  Alex Manoogian St. 1, 375025, Yerevan}
 \\
 $\;^2$ {\sl Artsakh State University, Mkhitar Gosh St. 3, Stepanakert}
 \\
 $\;^3$
 {\sl   Yerevan Physics Institute,  Alikhanian St. 2, 375036, Yerevan, Armenia

  }
%
\begin{abstract}
We give a simple geometric explanation for the similarity
transformation mapping one-dimensional conformal
mechanics to free-particle system.
Namely, we show that this transformation corresponds to the inversion
of the Klein model of Lobachevsky space (non-compact complex
projective plane) ${\widetilde{\DC P}}^1$. We also extend this
picture to the $\cN =2k$ superconformal mechanics described in
terms of Lobachevsky superspace ${\widetilde{\DC P}}^{1|k}$.
\end{abstract}

\subsection*{Introduction}
Conformal symmetry plays an important  role in modern field
theory. So, the study of various aspects of simple
(super)conformal invariant models could be useful for more complicated
systems. Since the middle of seventies, after \cite{fubini},
it was realized that even the one-dimensional
one-particle mechanics  given by the Hamiltonian
 \be
   H =\frac{{p}^2}{2}+\frac{g^2}{2x^2}
\label{h} \ee
 is a good polygon for the study of the possible consequences of the conformal symmetry.
 This Hamiltonian together with the generators
\be
D=px,\qquad  K=\frac{x^2}{2} \label{dk}
\ee
 forms the
conformal algebra $so(1,2)$ with respect to canonical Poisson
brackets $\{p, x\}=1$:
\be
\{  H , D\}= H  ,\qquad\{  H ,
K\}=2D,\qquad  \{ K, D\}= K.
\ee
Here $D$ is the dilatation, and
$K$ is the conformal boost. By that reason,  this system is called in literature
a one-dimensional "conformal mechanics".
Clearly, it is a conformal symmetric system in the
field-theoretical context (i.e. its action functional possesses
a conformal symmetry provided that the time reparametrizations are admitted)
but not in the sense of integrable mechanical systems.

Although this model is quite simple, it inherits some properties of more complicated
conformally  symmetric mechanical systems. The study of its various supersymmetric
 and superconformal extensions is of special importance. It was initiated  in \cite{scqm},
 and  has been continuing in the various directions up to now (see, e.g., \cite{iv,IKL} and references therein).
Let us mention the article \cite{kallosh} (and the related ones \cite{bhgalaj}), where  it was observed that the motion
of the (super)particle
near horizon of the extremal Raissner-Nordstr\"om black hole is described by the  (super)conformal mechanics.

The conformal mechanics can be considered as a two-particle Calogero model, which
 is a one-dimensional multi-particle integrable system with inverse-square interaction \cite{calogero}.
This model has attracted much attention due to numerous applications in the wide area of physics,
as well as due to rich internal structure (see, e.g., the recent review \cite{polychronakos} and references therein).
 Already in the pioneering paper \cite{calogero} it was observed that the spectrum of the Calogero model with
additional oscillator potential is  similar to the spectrum of
free $N-$dimensional oscillator. It was claimed there that a similarity transformation to free oscillator
system may exist, at least, in the part of  Hilbert space.
However,  this transformation has been written explicitly only decades  after \cite{hindu}.
Its elegant group-theoretical explanation  has been given  in  \cite{galajinsky}, where
the similarity transformation is related to the conformal group $SU(1,1)$.
Exploiting this similarity, the authors built not only ${\cal N}=2$
 supersymmetric Calogero model \cite{n2cal}
 but also suggested an algorithm for the construction of ${\cal N}=4$ superconformal
Calogero model, which was unknown. An explicit expressions for the ${\cal N}=4$
superconformal  Calogero system were presented in \cite{galaj4}.

In this  note we present a simple  geometric view on this "decoupling" transformation for the conformal mechanics
(i.e. for two-particle Calogero system) and for its ${\cal N}=2k$ superconformal extension.
First, we parameterize the phase space of conformal mechanics
by the Klein model of Lobachevsky space (which is a K\"ahler space) in such a way that the generators (\ref{h}), (\ref{dk})
become the isometries of the K\"ahler structure of the Klein model.
Then we show that the decoupling transformation corresponds to the inversion transformation
of the Klein model. The quantum counterpart of this picture can be get by the standard procedure
of the geometric quantization.

Then, using the above picture, we construct the ${\cal N}=2k$ superconformal mechanics
and give a similar description for its decoupling transformation as well.
For this purpose, we consider a linear action of the $u(1,1|k)$ algebra on the Euclidean superspace
$\DC^{1,1|k}$.
Then, performing the Hamiltonian reduction of this (phase) space by the $U(1)$ group, we arrive at
the Lobachevsky superspace ${\widetilde\DC P}^{(1|k)}$. The  superconformal algebra $su(1,1|k)$
defines the isometries of its K\"ahler structure. The Hamiltonian of the superconformal mechanics
 and the generators of superconformal algebra play the role
of the Killing potentials. The construction of the superconformal mechanics
by the reduction from the Euclidean superspace allows
us immediately extend the decoupling procedure to the supersymmetric system as well.

\subsection*{Conformal mechanics}
Let us start from the description of the conformal mechanics in a form,
which is suitable for our purposes. It is convenient to describe
the algebra \eqref{h} in terms of the generators
\be
J_0= H + K, \quad J_1=D, \quad
 J_3= H - K:
 \quad \{ J_a, J_b\}=-2\varepsilon_{abc} J^c,
\quad a,b,c=0,1,3, \label{u} \ee
where the indices  are upped  by
the use of $(1+2)$-dimensional Euclidean metrics $\gamma_{ab}=diag
(1,-1,-1)$. It is easy to see that these generators define the upper sheet of two-sheet hyperboloid
(Lobachevsky plane) with the radius $g$\footnote{We were informed by E.~Ivanov, that this had
been observed many years ago
in his old paper with Krivonos and Leviant \cite{per}.}:
\be
J_a J^a= g^2.
\ee
It can be parameterized by the complex coordinate
\be
{ w}=\frac{p}{x}+\frac{\imath g}{x^2}, \quad
{\rm Im}\;{ w}>0: \qquad \{{ w},\bar{ w}\}=-\frac{\imath}{g}\left({ w}-\bar { w}
\right)^2. \label{px}
\ee
So, the phase space of the system is the Klein model of the Lobachevsky plane.

In the above  parametrization, the $so(1,2)$ generators (\ref{u}) take the form
\be
 J_0=\imath g\frac{{ w}\bar { w} +1}{{ w}-\bar { w}},
\qquad
J_1=\imath g\frac{{ w}+\bar { w}}{{ w}-\bar { w}},
\qquad
J_3=\imath g\frac{{ w}\bar { w} -1}{{ w}-\bar { w}}.
\label{uk}
\ee
They are precisely the Hamiltonian
generators defining the isometries of the Kahler structure of the
Lobachevsky plane (Killing potentials), which is given by the
following metric and potential
\be
ds^2=-\frac{g d{ w} d\bar { w}}{({\bar w}- { w})^2},\qquad
\mathcal{K}=g\log \imath ({\bar w}- { w}). \label{klein}
\ee
This K\"ahler structure is invariant under the
 discrete transformation (which is, obviously, a
canonical one)
\be
{ w}\to -\frac{1}{  w}, \label{simt}
\ee
whereas the Killing potentials (\ref{uk}) are not invariant with
respect to this transformation:
\be
 J_0\to  J_0,\quad  J_1\to - J_1, \quad  J_3\to - J_3.
\label{simc}\ee
Taking into account (\ref{u}), one can rewrite this transformation
in terms of initial generators:
\be
 H =\frac{{g w}\bar { w}}{\imath ({\bar w}- {w})}\;
\to\;   K=\frac{g}{\imath ({\bar w}- { w})},\qquad K \to
-H,\qquad D\to -D. \label{h0}
\ee
So, the presented canonical
transformation maps the conformal mechanics to the free particle
system in the momentum space. Let us write down its explicit form  in $p, x$ coordinates:
\be
\frac{\tilde x}{\tilde p}+\imath \frac{g}{{\tilde
p}^2}=\frac{-px+\imath g}{p^2+ g^2/x^2}. \label{can}
\ee
Hence, we
get the following canonical transformation of the initial phase
space
\be {\widetilde p}=\pm\sqrt{{p^2}+{g^2}/{x^2}},\qquad
{\widetilde x}= \mp \frac{px}{\sqrt{{p^2}+{g^2}/{x^2}}}.
\label{can1}
\ee

{\bf Remark.}
One can  consider the quantum counterpart of our picture
performing the geometric quantization of $su(1,1)$ on Lobachevsky space.
The relevant formulae can be found in Ref. \cite{plyuschay}, where  the quantization was done
on the framework of Poincar\'e model of Lobachevsky space with the following  metric and
Killing potentials
\be
ds^2=\frac{g d{ w} d\bar { w}}{(1-{ w}\bar { w})^2},\quad   J_0=g\frac{1+{ w}\bar { w}}{1-{ w}\bar { w}},
\quad  J_3+\imath  J_1=\frac{2g\bar { w}}{1-{ w}\bar { w}},     \quad |{ w}|<1 .
\label{pmet}
\ee
Performing the conformal transformation
\be
{ w}\to\frac{{ w}-\imath}{{ w}+\imath}
\label{poin}\ee
we will arrive at the  Klein model of Lobachevsky space.

\subsection*{ Superconformal mechanics}
Our construction can be  extended straightforwardly to the
${\cal N}=2k$ superconformal algebra  $su(1,1|k)$
with some real central charge.
Apart from the $so(1,2)$ generators, this  superconformal algebra
contains $2k$ pairs  of Grassmanian odd generators $\Theta_\alpha^{\bar A}=(Q^{ A}, S^{ A})$,
$\overline\Theta_\alpha^A =({\overline Q}^A,{\overline  S}^A)$, $\alpha=1,2$,
$A=1,2,\ldots, k$, which are Hermitian
conjugates of each other, and the $u(k)$ generators $R^{AB}$.
Notice, that $Q^A$ and $\bar Q^A$ define the  supersymmetry generators,
and   $S^A$ and ${\overline S}^A$ the superconformal ones.

In order to define this algebra on the superextension of the
Lobachevsky space, let
us first consider  the $\DC^{1,1|k}$ Euclidean superspace
equipped  with the canonical K\"ahler (and symplectic)   two-form
\be
\imath (dz_1 \wedge d{\bar z}_0- dz_0 \wedge d{\bar z}_1) +d\eta^A \wedge d\bar\eta^A.
\label{2kahler}
\ee
The Poisson brackets  are given by  the following  non-vanishing
relations and their complex conjugates\footnote{
Let us remind, that K\"ahler
(super)metric is defined by the expression
$$ ds^2=dZ^A g_{A
{\bar B}}d{\bar Z}^B,\quad { \rm where}\quad
g_{A {\bar B}}=  \frac{\partial ^L}{\partial Z^A}
        \frac{\partial ^R}{\partial {\bar Z}^B}
                   K  (Z,{\bar Z}).$$
The Poisson brackets associated with this super-K\"ahler structure looks as follows
$$
\{ f,g\}=i\left( \frac{\partial ^R f}{\partial \bar z^A}
               g^{{\bar A}B}
            \frac{\partial  ^L g}{\partial z^B}
                   -
             \frac{\partial  ^R f}{\partial z^A}
             g^{ BA}
         \frac{\partial ^L g }{\partial \bar z^B}
               \right)
,\quad {\rm  where}\quad
g^{AB}g_{B C}=\delta^{ A}_{C}
\;\;,\;\;\;\; \overline{g^{ AB}}
 = (-1)^{p(A)p(B)}g^{BA}.$$ }

\be \{z_0,\bar z_1\}=1, \quad \{z_1,\bar z_0\}=-1,\quad
\{\eta^A,\bar\eta^B\}=\delta^{AB}.
\ee
 The rotational symmetries  of this
K\"ahler structure  defined by the following Killing potentials
\be
{\cal J}=\imath(z_1\bar z_0-z_0\bar z_1) +\imath\eta\bar\eta,
\label{J}
\ee
\be
{\cal J}_a=z\sigma^a\bar z, \quad R^{A\bar
B}=\imath\eta^A\bar\eta^B,\qquad \Theta_{\alpha}^{A}= {\bar
z}_\alpha\eta^A,
\label{unru}
\ee
which form the $u(1,1|k)$ superalgebra:
\be
\{ {\cal  J}_a,{\cal  J}_b\}=-2\varepsilon_{abc}
{\cal  J}^c,\quad
\{\Theta_{\alpha}^A, {\overline{\Theta_{\beta}^B}}\}=
\frac12\delta^{AB}\left(\sigma^a_{\alpha\beta} {\cal  J}_a +
\imath\epsilon_{\alpha\beta}{\cal J}\right) +
\imath\epsilon_{\alpha\beta}\left( R^{A B}-\frac 12 \delta^{AB}R_0
\right ),
\nonumber
\ee
\be
\{ {\cal J}_a,\Theta_{\alpha}^A\}=-\epsilon_{\alpha\beta}\sigma^a_{\beta\gamma}\Theta_{\gamma}^A,\quad
\{R^{A B},\Theta_{\alpha}^C\}=
\imath\delta^{C B}{\Theta}_{\alpha}^A,
\qquad
\{R^{A B}, R^{C D}\}=
\imath\delta^{C B}R^{A D}-\imath\delta^{A D}R^{C B},\label{un111}
\ee
\be
\{{\cal J},  {\cal  J}_a\}=\{{\cal J}, \Theta_\alpha^A\}=\{{\cal J},
R^{A B}\}=0,\qquad \{{\cal J}_a,
R^{A B}\}=0, \label{center}
\ee
where
$\sigma^a=\sigma^0, \sigma^1, \sigma^3$
are "Minkowskian" Pauli matrixes ($\sigma^0$ is the identity matrix)
and $R_0=\sum_A R^{A A}$. The generator
${\cal J}$ defines  the center of the superalgebra $u(1,1|k)$,
while the other generators form the superconformal
algebra $su(1,1|k)$. Hence, we can reduce the $\DC^{1,1|k}$
superspace by the action of the generator ${\cal J}$
to the Lobachevsky superspace, whose isometry superalgebraalgebra is  $su(1,1|k)$.
We refer to the \cite{khud} for the details, where the complex projective
superspace $\DC P^{n|k}$ has been constructed by the Hamiltonian
reduction from the Euclidean space $\DC^{n+1|k}$.
The procedure presented below is just its  noncompact counterpart.
The details of reduction of  $\DC^{1,1}$ to the Poincare and Klein
models can be found in \cite{anyon}.
Clearly, the reduced superspace is $(2|2k)_{\DR}$-dimensional.
One can parameterize it by
\be
\breve{w}=\frac{z_0}{z_1},\qquad \theta^A=\frac{\eta^A}{z_1} :
\qquad\{{\breve{w}},{\cal J}\}=\{\theta^A ,{\cal  J}\}=0.
\label{redcor}
\ee
On the level surface $\mathcal{J}=g$ the following relation yields:
\be
|z_1|^2=\frac{g}{\imath(\bar{ \breve{w}}-{ \breve{w}})+\imath\theta\bar\theta}.
\label{z0}
\ee
Calculating the Poisson brackets between coordinates (\ref{redcor})
and restricting  them to the level surface ${\cal J}=g$,
we obtain the Poisson brackets on the reduced phase space:
\be
\{\breve{w},\bar{ \breve{w}}\}=\frac{\imath (\bar{ \breve{w}}-{\breve{w}})+\imath\theta\bar\theta}{g}(\breve{w}-\bar{ \breve{w}}),\qquad
\{\breve{w},\bar\theta^A\}=-\imath \frac{\imath ( \bar{ \breve{w}}-{\breve{w}})+\imath\theta\bar\theta}{g}\bar\theta^A,
\qquad \{\theta^A,\bar\theta^B\}=\frac{\imath (\bar{ \breve{w}}-{\breve{w}})+\imath\theta\bar\theta}{g}\delta^{AB}
\label{suppois}\ee
These Poisson brackets define the K\"ahler structure given by  the
following K\"ahler potential:
\be
{\widetilde{\mathcal{K}}}=g\log (\imath(\bar{ \breve{w}}-{\breve{w}})
+\imath\theta\bar\theta). \label{skah}\ee
This supermanifold  is a superextension of the Klein model of the Lobachevsky space, $\widetilde{\DC P}^{1|k}$.

Now, let us write down the Killing potentials of the $su(1,1|k)$ superalgebra
obtained by the restriction of
(\ref{unru}) to the level surface ${\cal J}=g$
\be
 {\cal  J}_0={\cal H}+{\cal K} =g\frac{\breve{w}\bar{\breve{w}} +1}{\imath(\bar{ \breve{w}}-{ \breve{w}})+\imath\theta\bar\theta},
\quad
 {\cal  J}_1={\cal D}= g\frac{\breve{w}\bar{\breve{w}} -1}{\imath(\bar{ \breve{w}}-{ \breve{w}})+\imath\theta\bar\theta}
,\quad
 {\cal  J}_3={\cal H}-{\cal K}= g\frac{\breve{w}+\bar{\breve{w}}}{\imath(\bar{ \breve{w}}-{ \breve{w}})+\imath\theta\bar\theta },
\label{su111gen}\ee
\be
\Theta_1^A=Q^A=\frac{g\bar\theta^A}{\imath(\bar{ \breve{w}}-{ \breve{w}})+\imath\theta\bar\theta}
,\qquad \Theta_2^A=S^A=
g\frac{ \breve{w}\bar\theta^A}{\imath(\bar{ \breve{w}}-{ \breve{w}})+\imath\theta\bar\theta},
\qquad R^{A\bar B}=g\frac{\imath \theta^A\bar\theta^{B}}{\imath(\bar{ \breve{w}}-{ \breve{w}})+\imath\theta\bar\theta}.
\label{such}\ee
These generators form the algebra (\ref{un111}) with respect to the reduced Poisson brackets, where
 the generator $\mathcal{J}$ is replaced by the cental charge $g$.
It is clear from our consideration that the supersymmetric extension
of the similarity transformation looks as follows
\be
\breve{w}\to -\frac{1}{\breve{w}},\qquad \theta^A\to -\frac{\theta^A}{\breve{w}}.
\label{susim}\ee
It yields the following transformation of the generators of $\cN =2k$ superconformal algebra
\be
{\cal H}\to {\cal K},\quad {\cal K}\to {\cal H},\quad {\cal D}\to -{\cal  D},\quad  R^{A B}\to R^{A B},\quad  Q^A\to S^A,
\quad S^A\to Q^A.
\label{sutrans}\ee
Finally, we pass to the coordinates, which split
the ferminic and bosonic sectors of the Poisson brackets:
\be
\chi^A=\frac{\sqrt{g}\theta^A}{\sqrt{\imath(\bar{ \breve{w}}-{\breve{w}})+\imath\theta\bar\theta}}\;,
\quad {w}=\breve{w}-\frac 12 \theta\bar\theta:
\ee
\be
\{{w},\bar { w}\}=-\imath\frac{({  w}-{\bar  w})^2}{g},\quad \{\chi^A,\bar\chi^B\}=\delta^{AB}, \quad
\{w,\bar\chi^A\}=\{w,\chi^B\}=0.
\label{sukan}\ee
Substituting these expressions in (\ref{su111gen}), (\ref{such})  and expressing
${\tilde w}$ via canonical coordinates $(p, x)$  as in (\ref{px}),
we get the $\cN=2k$ superconformal mechanics.
In the same manner as in the pure bosonic case, we obtain the expression of the similarity
transformations (\ref{susim}).

\subsection*{Summary and discussion}
In conclusion, let us emphasize the main statements of the
current article.
\begin{itemize}
\item
We identified the Killing potentials of the Klein model of the
Lobachevsky space with the Hamiltonian of classical  one-dimensional conformal mechanics
and with the generators of conformal boost and dilatation.
The inversion transformation (with minus sign) corresponds to the  canonical transformation
of the  Hamiltonian to the generator of conformal boost, which describes
the one-dimensional free particle.
In other words, the inversion transformation of the Lobachevsky space
defines the "decoupling transformation" of conformal mechanics.
\item
Using the method of Hamiltonian reduction, we
constructed the one-dimensional ${\cal N}=2k$ superconformal mechanics. The generators of
its dynamical symmetry superalgebra $su(1,1|k)$ define the Killing potentials of the
superextension of the Klein model of the Lobachevsky space corresponding
 to the noncompact complex projective superplane $\widetilde{\DC P}^{1|k}$.
We found the decoupling transformation for this superconformal system as well.
\end{itemize}
It seems that the presented picture
can be extended without much efforts to
higher-dimensional conformal mechanics and  Calogero model. It would be interesting to find a similar description for
systems with less trivial potentials, and  for the
higher-dimensional superconformal mechanics with the
Dirac monopole  (the  example of such system was
suggested in \cite{IKL}). Particularly, we expect that noncompact complex projective space $\DC P^N$ can  be related
with integrable multi-particle systems.

  Note  that  the dilatation operator,
providing the phase space of the superconformal mechanics with a constant curvature,
does not play  any role in our consideration.
Hence, one can suppose that a similar construction may be realized  for the specific case
of two-dimensional surfaces like {\sl the Special K\"ahler surfaces of the local type}  with two isometries.

Finally, let us draw attention to the pretty simple form of the superconformal mechanics
given by the Poisson brackets (\ref{suppois}) and the generators (\ref{su111gen}), (\ref{such}).
The simplicity is due to non-canonical Poisson bracket with obvious geometrical meaning, which we
used instead of canonical ones. In fact, the quite transparent geometrical nature of
these expressions, as well as the way, via which they were found (Hamiltonian reduction),
allow us to believe that this could be the shortest and most natural way  for the construction of the $\cN \geq 4$
superconformal Calogero model.
\\

{\large Acknowledgments.} We are grateful to Anton Galajinsky for the useful comments  and encourage, and
 to Segey Krivonos and  Vadim Ohanyan for the interest in work. We acknowledge Evgeny Ivanov and Mikhail Plyushchay for pointing
 out to their papers related to the subject.
The work is supported by the grant of Artsakh Ministry of Education and Science and
by the    NFSAT-CRDF UC 06/07 and  INTAS-05-7928 grants.

\end{document}